# TURNING MUSIC IDENTIFICATION INTO A NEURAL FORWARD PASS

*Muhammad Taimoor Haseeb*,†, Ahmad Hammoudeh*,†, Gus Xia*
*Music X Lab, Mohamed Bin Zayed University of Artificial Intelligence, UAE*

** Corresponding authors: Muhammad Taimoor Haseeb, Ahmad Hammoudeh {Muhammad.Haseeb, Ahmad.Hammoudeh}@mbzuai.ac.ae*
*† These authors contributed equally.*

**Abstract**

*Search, a foundational operation in computer science, maps a query to a matching item in a collection. It is typically implemented as a System-2 like, rule-based pipeline in which a key is computed, an index is probed, and candidates are verified. By contrast, human recognition resembles a System-1 like, associative model of identity recovery, in which even partial cues can trigger a recall without explicitly enumerating, ranking, or even accessing discrete candidates. Here, we show that music sound identification, a difficult search problem, can be performed in a single neural feed-forward pass by a generative transformer. Trained on an audio dataset, the model predicts the corresponding track identifier from a short audio excerpt. This approach surpasses state-of-the-art acoustic fingerprinting, with the largest gains for short audio segments (1 second), demonstrating the method is not only viable but advantageous. Moreover, it reduces external storage to 0.33% of the baseline footprint and improves inference latency by 2.3× (p95). Furthermore, the model can reject queries for unseen tracks, supporting open-set operation while reducing misattribution risk. Using music track identification as an example, this work reframes search, bringing it closer in spirit to human associative recognition and away from algorithmic database lookup.*

**Main**

Search is a foundational operation in computer science that maps a query to a matching item in a collection. This deceptively simple operation has shaped early algorithmic theory, data structure design, and database architecture. In practice, search is treated as a System 2 operation: an explicitly engineered, rule-based pipeline that computes a representation of the query, probes an index built over the stored collection, and verifies candidates [19]. Decades of innovative optimizations — such as indexing, caching and sharding — have pushed this design to extraordinary scale, making it possible to query a retailer's inventory, the documents, code, or logs that run modern organizations, and even the web. While these optimizations are an engineering triumph, and undoubtedly make search faster and more efficient, the underlying procedure remains fundamentally a System 2 process.

By contrast, human recognition suggests an associative, cue-driven model of identity recovery rather than an explicit search over enumerated candidates. Partial cues trigger a System-1-like recall, where the answer arrives before deliberation even begins — just as a melody announces itself within the first few notes. In closed-set identification, a query is not a request to discover something new, but an incomplete, sometimes corrupted view of a single item, drawn from a fixed collection, and the task is simply to recover its identity. In that sense, human cognition appears to solve this search problem through direct recognition rather than by explicitly enumerating or ranking discrete candidates. In this study, we recast a traditionally System-2 search problem as a System-1-like recognition task using Generation Augmented Retrieval (GAR); bringing it closer in spirit to human associative recognition and away from algorithmic database lookup.

Specifically, we show that a music track identification system can be implemented without database lookup at inference. The dominant approach, acoustic fingerprinting, solves for track identification by constructing salient time–frequency event pairs, hashing these fingerprints, and matching queries against their external database [1, 2]. Newer learning-based methods improve fingerprint quality, but still depend on explicit retrieval and ranking of candidates [3]. Instead, GAR uses a generative decoder to predict a discrete track identifier, encoded as a short token sequence, directly from an audio query. This shifts the reference collection from an external database to instead be stored implicitly in the model parameters, allowing identity to be recovered in a single neural forward pass.

Music track identification serves as an ideal testbed for direct recognition for several reasons. First, it offers a particularly sharp test of direct recognition as each audio query is an observation of a single track from a dataset, and success is measured by predicting the exact identifier rather than a ranked list. Second, audio queries are unusually harsh in practice, arriving at arbitrary time offsets, and undergoing routine transformations from background interference, reposting artifacts, compression, etc., thereby providing a stringent test for direct recognition. Third, fingerprinting has been refined over decades into a mature, purpose-built retrieval system, providing an excellent comparative baseline for an approach that seeks to dispense with explicit retrieval. Together, these make track detection both conceptually clean and empirically stringent.

Our results demonstrate that direct recognition is not only viable but advantageous. Using this testbed, we show that direct recognition surpasses fingerprinting, with the largest gains for short 1-second audio segments where conventional methods are most brittle. Moreover, direct recognition reduces external storage to 0.33% of the baseline footprint while improving latency by 2.3× (p95). We further analyze the scaling behavior of our approach across varying dataset sizes, extend our method to open-set operation with explicit out-of-dataset rejection, and examine continual-learning strategies for maintaining an evolving dataset.

## Results

We evaluate direct recognition for music track identification against two retrieval-based baselines. The first is Dejavu, a widely adopted open-source acoustic fingerprinting system that detects spectrogram peaks, hashes peak pairs into fingerprints, and matches those fingerprints against a database [15]. The second is GraFPrint [20], a recent GNN-based audio identification method that learns neural fingerprints while still performing retrieval against a reference database at inference [20]. Details including dataset construction, training protocols, query length, etc., are provided in Methods. Unless otherwise stated, experiments use a dataset of $N$ = 1,000 tracks, queries sampled at random temporal offsets, and Top-1 identification accuracy defined as exact ID match.Each experiment is repeated over five pre-specified random seeds, and results are reported as the mean across the five runs.

**Table 1: Key performance results summary** *(Query Length (ℓ) = 1s, Dataset Size (N) = 1,000 tracks)*

| | Closed-set identification | | Open-set rejection (in-dataset vs out-of-dataset queries) | | | System | | |
|---|---|---|---|---|---|---|---|---|
| | **Top 1 Accuracy** (%, ↑) (clean FMA) | **Top 1 Accuracy** (%, ↑) (noisy dataset) | **False Positive Rate** (%, ↓) | **False Negative Rate** (%, ↓) | **Balanced Accuracy** (%, ↑) | **Inference Latency (CPU)** (95th pct., ms/query, ↓) | **Inference Latency (GPU)** (95th pct., ms/query, ↓) | **Database Storage** (MB,↓) |
| **Direct Recognition** | **99.6** | **93.3** | 2.4 | **4.8** | **96.4** | **97** | **93** | **10** |
| **Dejavu** | 76.3 | 37.2 | **1.2** | 30.4 | 84.2 | 224 | N/A | 3,348 |
| **GraFPrint** | 86.6 | 51.2 | 3.0 | 7.8 | 94.6 | 479 | 351 | 1,182 |

Direct recognition changes the overall operating regime of music identification. Table 1 summarizes key results across all experiments. Direct recognition is more accurate in closed-set identification, more robust under noise, achieves a better open-set trade-off, and does so with lower latency and dramatically smaller external storage. In the following sections, we examine these differences in greater detail.

### Impact of Query Length on Identification Accuracy

We benchmark direct recognition as a function of query length. Queries are sampled at random offsets and evaluated under clean digital conditions. Table 2 below reports Top 1 accuracy across ℓ ∈ {1,2,3,5,10} query length and compares against fingerprinting baselines.

**Table 2: Top 1 Accuracy (%, ↑) across varying query lengths (ℓ)**

| | **ℓ=1s** | **2s** | **3s** | **5s** | **10s** | **Notes** |
|---|---|---|---|---|---|---|
| **Direct Recognition** | **99.6** | **99.8** | **99.9** | **99.0** | 99.5 | *ℓ ∈ {1s, 2s, 3s, 5s, 10s} query length*<br>*N = 1,000 tracks*<br>*Clean audio signal*<br>*Closed-set (in-dataset)*<br>*Model size ∈ {Small (S)}* |
| **Dejavu** | 76.3 | 92.2 | 96.6 | 98.7 | 99.5 | |
| **GraFPrint** | 86.6 | 91.2 | 92.9 | 94.2 | 95.5 | |

Direct recognition consistently achieves high accuracy across all query lengths, but is most consequential when queries are very short, and where retrieval pipelines are most constrained. This weak dependence on duration suggests that the model can recover identity from sparse, partial evidence. Retrieval-based baselines show the opposite pattern with strong improvement on longer excerpts, indicating that short-query performance gap is primarily an information problem.

### Impact of Dataset Size and Model Capacity on Identification Accuracy

To evaluate how direct recognition scales as the search space grows, we vary dataset size $N \in \{100, 1,000, 5,000, 10,000, 25,000\}$. Because this behavior may depend on model capacity, we also evaluate Small, Medium, and Large model variants. Table 3 reports Top 1 accuracy at $\ell \in \{1s, 5s\}$.

**Table 3: Top 1 Accuracy (%, ↑) across increasing dataset size (*N*).**

| | $\ell$ | N=100 | 1,000 | 5,000 | 10,000 | 25,000 | Notes |
|---|---|---|---|---|---|---|---|
| **Direct Recognition** *(Small Model)* | 1s | 99.6 | **99.6** | **98.2** | 89.6 | 77.9 | *N ∈{100, 1,000, 5,000, 10,000, 25,000} tracks*<br>*ℓ ∈{1 s, 5s} query lengths*<br>*Clean audio signal*<br>*Closed-set (in-dataset)*<br>*Model size ∈{Small (S), Medium (M), Large (L)}* |
| **Direct Recognition** *(Medium Model)* | 1s | 99.7 | 99.2 | 98.1 | 97.3 | 95.3 | |
| **Direct Recognition** *(Large Model)* | 1s | **100.0** | 98.7 | 98.1 | **97.8** | **96.9** | |
| **Dejavu** | 1s | 73.3 | 76.3 | 70.9 | 68.1 | 61.9 | |
| **GraFPrint** | 1s | 96.0 | 86.6 | 80.8 | 78.1 | 74.5 | |
| **Direct Recognition** *(Small Model)* | 5s | **100.0** | 99.0 | 97.3 | 93.3 | 86.5 | |
| **Direct Recognition** *(Medium Model)* | 5s | 100.0 | 99.6 | 99.5 | 97.5 | 94.5 | |
| **Direct Recognition** *(Large Model)* | 5s | 100.0 | **99.7** | **99.9** | **97.8** | **97.1** | |
| **Dejavu** | 5s | 97.8 | 98.7 | 98.4 | 97.6 | 95.3 | |
| **GraFPrint** | 5s | 97.8 | 94.2 | 91.1 | 90.0 | 88.8 | |

As the search space grows, direct recognition degrades gracefully. Up to $N$=5,000, all three model sizes remain near 98–100% Top-1 accuracy. At $N$=25,000 in the 1s setting, the small model drops to 77.9%, while the medium and large models achieve 95.3% and 96.9% respectively, remaining far above both retrieval baselines. Longer queries provide more discriminative evidence, making the task easier for all methods. GraFPrint declines more rapidly with dataset size, whereas Dejavu becomes competitive mainly when longer excerpts provide sufficient information. Moving from the small to the medium model recovers most of the scaling loss, with only marginal additional benefit from further scaling. Medium and large direct-recognition models remain near ceiling across all $N$.

### Impact of Noise and Version Shift on Identification Accuracy

Next, we stress-test robustness to identity-preserving variation induced by local waveform corruption and version shifts. Noise denotes additive background interference applied to the original recording, while radio edits and live versions denote alternate versions of the same musical work that may differ structurally, acoustically, or in performance. Details of this curated dataset are provided in Methods under "Curated version-shift robustness set." Table 4 summarizes results across these conditions. Clean results in this table are evaluated on the original-reference recordings from the curated version-shift set, not on the FMA split used in Tables 2–3.

**Table 4: Top 1 Accuracy (%, ↑) under noise and version-shift conditions**

| | $\ell$ | Clean | Noise | Radio Edit | Live Version | Notes |
|---|---|---|---|---|---|---|
| **Direct Recognition** | 1 s | **99.6** | **93.3** | 85.5 | 24.9 | *N = 1,000 music tracks (each with paired original, noisy, radio-edit, and live-version recordings).*<br>*ℓ ∈{1 s, 5s} query lengths*<br>*Clean, additive-noise, radio-edit, and live-version queries*<br>*Closed-set (in-dataset)* |
| **Dejavu** | 1 s | 77.1 | 37.2 | 54.9 | 8.4 | |
| **GraFPrint** | 1 s | 96.0 | 51.2 | **89.8** | **29.8** | |
| **Direct Recognition** | 5s | **100.0** | **91.0** | 89.2 | 28.3 | |
| **Dejavu** | 5s | 99.3 | 56.9 | 84.9 | 15.7 | |
| **GraFPrint** | 5s | 97.8 | 63.2 | **94.4** | **53.4** | |

Real-world distortions reveal a clear split between robustness to waveform corruption and robustness to version shift. Direct recognition consistently outperforms Dejavu across all conditions, with the largest gains under additive noise, where it remains strong even for 1-s queries. Relative to GraFPrint, direct recognition is strongest on clean and noisy queries, indicating better robustness to local corruption, whereas GraFPrint performs better on radio edits and live versions, especially with longer 5-s context. Live recordings are the hardest condition for all methods, and the sharp drop in accuracy suggests that large changes in acoustics, mixing, and performance substantially challenge exact ID recovery. Overall, direct recognition is particularly effective for short, noisy queries.

**Performance Under Open-Set Operation**

In deployment, many queries fall outside the monitored dataset. We therefore evaluate open-set rejection by augmenting the test set with out-of-dataset tracks and allowing systems to abstain when confidence is low [12]. Open-set performance is not only about rejecting unseen tracks, but about doing so without rejecting too many in-dataset queries. To this end, we use a multi-segment protocol in which each query is split into $k = 6$ non-overlapping 1s segments, whose predictions are aggregated by majority vote and accepted based on cross-segment agreement. Because fingerprinting doesn't work as well with short query lengths, we also repeat the experiment with a longer 6s query ($\ell = 6$ s, $k = 1$). Table 5 and Figure 1 summarize the results.

**Table 5: Open-set identification with multi-segment aggregation.**

| | $\ell$ | $k$ | F1 Score ↑ | Accuracy ↑ | Recall ↑ | Precision ↑ | Notes |
|---|---|---|---|---|---|---|---|
| **Direct Recognition** | 1 s | 6 | **96.3** | **96.4** | 95.2 | 97.5 | *N = 1,000 audio tracks*<br>*$\ell$ = 1 s query lengths*<br>*Clean audio signal*<br>*Open-set (out-of-dataset)*<br>*k ∈ {6} segments / query* |
| **Dejavu** | 1 s | 6 | 9.7 | 52.6 | 5.1 | 100.0 | |
| **GraFPrint** | 1 s | 6 | 94.5 | 94.6 | 92.2 | 96.9 | |
| **Dejavu** | 6 s | 1 | 81.5 | 84.2 | 69.6 | **98.3** | *$\ell$ = 6 s query lengths*<br>*k ∈ {1} segments / query* |
| **GraFPrint** | 6 s | 1 | 95.7 | 95.7 | **96.3** | 95.1 | |

**Figure 1: Confusion matrix comparing classification performance on openset identification**

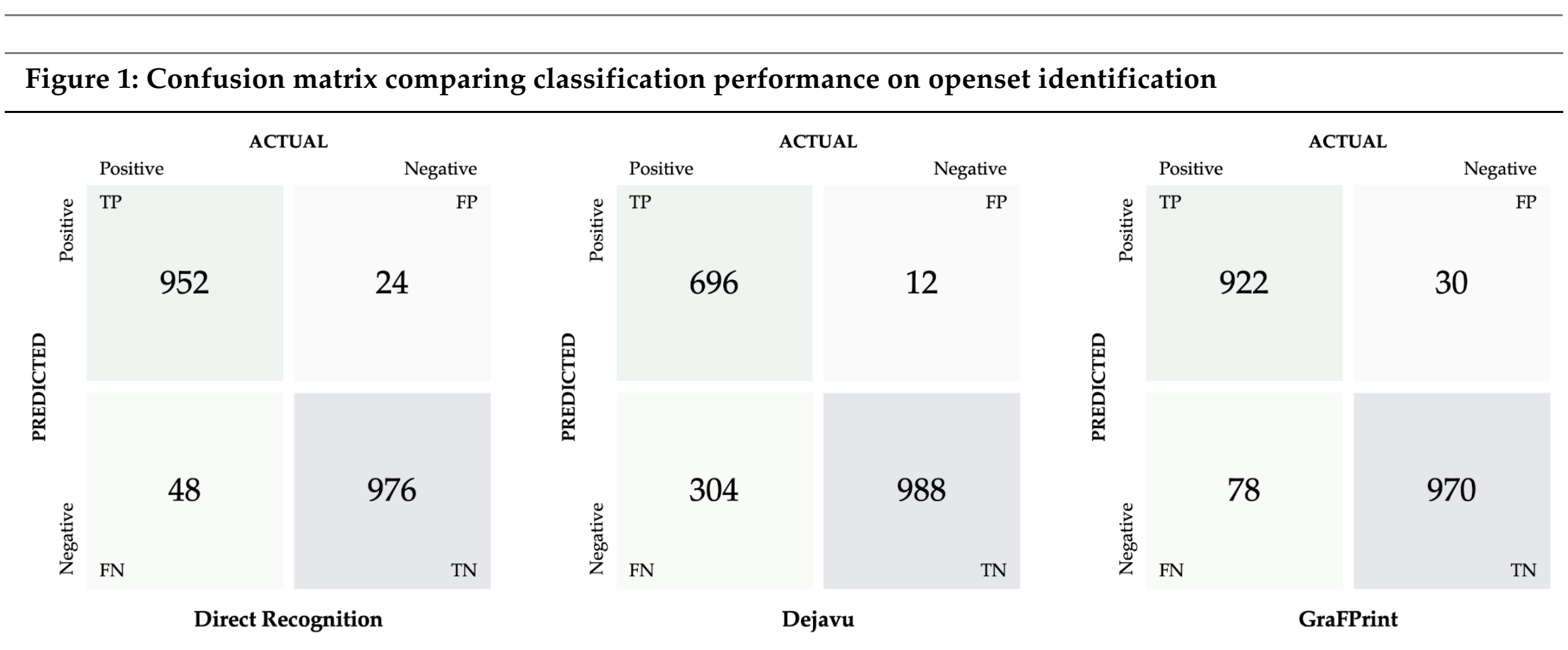

Under these settings, direct recognition gives the cleanest trade-off as it keeps both false accepts and false rejects low, rather than achieving safety by becoming overly conservative. GraFPrint is also strong, suggesting that learned representations help in the open-set regime, but direct recognition still makes the fewest total mistakes. Dejavu's high precision is obtained largely by missing many true in-dataset queries, and even with longer excerpts it remains recall-limited. Overall, agreement across short segments provides a useful confidence signal, making direct recognition reliable not only for identifying known tracks, but also for deciding when to abstain.

**Systems Performance and Storage Analysis**

We next evaluate the system implications of eliminating retrieval. Specifically, we compare inference latency and external storage requirements between direct recognition and fingerprinting baselines under identical hyperparameter configuration, hardware and dataset size. Retrieval-based systems rely on a per-track external index and database probing; whose latency depends on index size, layout, and memory traffic. In contrast, direct recognition performs a fixed neural forward pass. Table 6 summarizes latency and storage measurements.

**Table 6: External storage and latency performance comparison**

| | *Hardware* | **Inference Latency** (50th pct., ms/query, ↓) | **Inference Latency** (95th pct., ms/query ↓) | **Model size** (# parameters) | **External storage** (MB, ↓) | **Notes** |
|---|---|---|---|---|---|---|
| **Direct Recognition** *(Small Model)* | CPU | 72 | 97 | 2,764,800 | **10** | *N = 1,000 audio tracks*<br>*ℓ = 1 s query lengths*<br>*Clean audio signal*<br>*Open-set (out-of-dataset)*<br>*k = 1 segment / query*<br>*CPU: Intel Xeon @ 2.20 GHz*<br>*GPU: NVIDIA L4* |
| **Direct Recognition** *(Medium Model)* | CPU | **71** | **95** | 10,247,424 | 39 | |
| **Direct Recognition** *(Large Model)* | CPU | 94 | 119 | 19,310,976 | 74 | |
| **Dejavu** | CPU | 132 | 224 | - | 3,348 | |
| **GraFPrint** | CPU | 468 | 479 | 20,620,576 | 1,182 | |
| **Direct Recognition** *(Small Model)* | GPU | **70** | **93** | 2,764,800 | **10** | |
| **Direct Recognition** *(Medium Model)* | GPU | 70 | 94 | 10,247,424 | 39 | |
| **Direct Recognition** *(Large Model)* | GPU | 85 | 108 | 19,310,976 | 74 | |
| **Dejavu** | GPU | - | - | - | 3,348 | |
| **GraFPrint** | GPU | 347 | 351 | 20,620,576 | 1,182 | |

Eliminating retrieval yields a more predictable forward-pass based inference path. On an Intel Xeon @ 2.20 GHz CPU, Direct recognition is substantially faster than Dejavu, about 1.8× at median latency and about 2.3× at the 95th percentile; while requiring only 11–74 MB of storage versus 3.3 GB for Dejavu and 1.18 GB for GraFPrint. The small and medium models have nearly identical latency, suggesting that end-to-end runtime is not strongly sensitive to model size, though the larger model shows a marginal latency increase.

**Maintaining an Internalized Dataset Under Incremental Updates**

A practical systems question is how an internalized dataset can be maintained as new tracks are added. Unlike retrieval-based systems, which accommodate growth by inserting new fingerprints into an external index, direct recognition requires updating model parameters. We therefore evaluate several update strategies, including full fine-tuning, Learning without Forgetting (LwF), and rehearsal-based methods [11]. Specifically, we measure (i) retention on previously learned tracks, (ii) acquisition of newly added tracks, and (iii) overall identification accuracy after each update. Table 7 summarizes our results.

**Table 7: Maintaining an internalized dataset under incremental updates**

| | **Retain** Old dataset accuracy (%, ↑) | **Acquire** New dataset accuracy (%, ↑) | **Overall** Old + New dataset accuracy (%, ↑) | **Forgetting** Pre – Post update accuracy on old dataset (%, ↓) | **Cost** NVIDIA L4 GPU Training (Hour, ↓) | **Notes** |
|---|---|---|---|---|---|---|
| **Full retraining** | 97.0 | 96.6 | 96.8 | 2.6 | 127.5 | *Retrain on old + new dataset* |
| **Fine-tuning** (all weights) | 0.0 | **99.6** | 49.8 | 99.6 | **3.6** | *Fine-tune only on new dataset* |
| **Learning without Forgetting** | 57.8 | 17.8 | 37.8 | 41.8 | 21.8 | *Distillation to preserve old behaviour* |
| **Rehearsal** (50%) | **99.0** | 98.8 | 98.9 | **0.6** | 23.0 | *Fine-tune on new data with 50% old samples replayed / epoch.* |
| **Rehearsal** (20%) | 98.4 | **99.6** | **99.0** | 1.2 | 11.7 | *Fine-tune on new data with 20% old samples replayed / epoch.* |

Full retraining serves as a baseline, achieving 97.0% retention on old tracks and 96.6% acquisition on new tracks at a cost of 127.5 GPU-hours. At the opposite extreme, naïve fine-tuning on new data only attains 99.6% acquisition with just 3.6 GPU-hours, but catastrophically forgets the internalized dataset. Among parameter-efficient strategies, Learning without Forgetting (LwF) only partially mitigates this trade-off. It improves retention relative to naïve fine-tuning but fails to acquire the new dataset, yielding 37.8% overall accuracy at 21.8 GPU-hours. Rehearsal, which includes fine-tuning on new data while replaying a fraction of old samples each epoch, achieves the best balance. With 50% replay, the model reaches 99.0% retention and 98.8% acquisition at 23.0 GPU-hours. Even with 20% replay, acquisition remains 99.6% at 11.7 GPU-hours, with 1.2-point forgetting and 99.0% overall accuracy.

**Discussion**

Recent works have begun to narrow the gap between learned models and retrieval pipelines. Embedding-based audio-identification methods improve matching robustness, but identity is still resolved by nearest-neighbor, maximum-inner-product, or fingerprint search over an external reference database [3, 17, 18, 20]. These systems can be accelerated with approximate nearest-neighbor and GPU similarity-search infrastructure [8, 10]. Learned index structures go further by modelling parts of the indexing, yet the retrieved data remains in an external database during inference [4]. Generative retrieval in text predicts item identifiers directly, moving the index into model parameters [5, 6, 7]. However, modalities where both queries and identifiers are discrete, like text, are easier.

Unlike text, music queries are continuous and time-shifted, and are routinely transformed by compression, noise, mixing, or device capture. Music track identification makes this concrete, where systems must identify short excerpts within large datasets. There is little semantic ambiguity to hide behind as each query has a single correct identifier. By contrast, many text retrieval tasks admit multiple valid targets and depend on user intent and context, complicating the separation between dataset internalization and relevance modeling. Our scope is closed-set identification, where each query corresponds to exactly one dataset item and success is exact identifier recovery, rather than relevance-ranked retrieval. Music track identification provides a stringent and conceptually clean testbed for this setting.

We demonstrate that closed-set identification over a fixed audio dataset can be implemented as direct neural recognition. A decoder maps short audio excerpts directly to a tokenized track identifier by autoregressively generating track IDs, replacing the retrieve-and-verify pipeline. Across clean and degraded conditions, this approach surpasses Dejavu [15]. The largest gains are observed for one-second queries where conventional pipelines are most brittle [3, 16, 17, 18]. Because the dataset is internalized in model parameters rather than stored in an external table, direct recognition reduces storage to 0.33–2.21% of the Dejavu footprint, depending on model size. For a fixed architecture, inference latency largely becomes a function of compute, independent of dataset size. The model also supports open-set operation through a confidence signal from cross-segment agreement.

One way to view closed-set identification is as amortized maximum-a-posteriori (MAP) inference over a finite hypothesis class. Each track identity corresponds to a distribution over observable excerpts produced by time offset and nuisance transformations. Under 0-1 loss, Bayes-optimal decision maps an excerpt to the identity with the largest posterior probability: $arg\ max_y\, p(y \mid q)$. Direct recognition amortizes the inference into parameters. By training on diverse samples of each dataset track, the model learns internal representations and a parametric approximation to the posterior that aggregates weak cues into a concentrated identity decision. Our empirical results indicate that these learned representations match or exceed the invariances encoded by fingerprinting without consulting an external database.

Retrieval pipelines do not model $p(q|y)$ explicitly. Instead, they compute index representation of each dataset track $\varphi(q)$, and resolve queries through alignment. In fingerprinting systems, $\varphi(q)$ consists of sparse time–frequency event pairs that are hashed for efficient database lookup. This representation is designed to be invariant to common distortions such as noise, compression, and temporal offset. However, because $\varphi$ is highly compressed and non-invertible, it retains only part of the signal detail. For short queries, the number of observed fingerprints is small, limiting the identity information retained in $\varphi(q)$ and making closely related recordings difficult to distinguish. From this perspective, performance is constrained primarily by the informativeness of the representation rather than by the mechanics of retrieval.

This framing also clarifies the computational trade-off between internalization and retrieval. For a fixed architecture, decoder inference applies a fixed compute graph whose cost depends on query length but not on dataset size, since no per-track entries are consulted at inference. However, independence from $N$ in computation does not eliminate dependence on $N$ in information. Selecting one identity among $N$ requires $O(\log N)$ bits of information, which must be recovered from a nuisance transformed observation. As N grows relative to model capacity, identities become harder to separate and accuracy degrades, consistent with our empirical scaling results.

Therefore, the number of identity excerpt distributions that can be cleanly separated is constrained by model capacity. This is consistent with the scaling behavior observed in our experiments. This reasoning extends beyond music to any closed-set recognition problem in which offline optimization trades search for a forward pass.

In contrast, retrieval systems make a different trade-off. Classical fingerprinting stores explicit per-item fingerprints in an external database whose footprint grows roughly linearly with dataset size and total duration. During inference, the system extracts $H$ fingerprints from an audio excerpt, and resolves identity using $O(H)$ hash probes, plus any verification and alignment over the returned postings. Retrieval systems trade model capacity limits for index growth and query-time search, whose compute, memory footprint, and tail latency scale with dataset size and fingerprint density. At a systems level, retrieval relies on irregular accesses into a large external index, making performance sensitive to memory traffic and data layout. A forward pass, on the other hand, repeatedly applies the same fixed computation over a compact checkpoint with high locality. Once per-track lookup is removed, latency is largely determined by the cost of that fixed computation rather than by index size, which helps explain its stability as the dataset grows.

In most machine learning settings, memorization is treated as overfitting because the goal is to generalize beyond the training data. In direct recognition, storing the database *inside* the model is not a failure mode, it is the mechanism. The scientific question is therefore not whether the model memorizes, but what it must generalize over while doing so. For audio identification, the required generalization is within-identity. Specifically, robustness to offsets and transformations that preserve identity, while remaining sensitive enough to separate closely related recordings. This setting provides a tractable way to study how a model can store a fixed collection in its parameters yet generalize to the many ways those items appear in the wild.

While the discussion so far assumes a closed-set regime in which every query corresponds to a stored identity, real-world deployment must also address open-set operation as many queries fall outside the monitored dataset. Direct recognition derives a confidence signal from cross-segment agreement. With multiple short segments, cross-segment agreement supports acceptance and disagreement supports rejection. This produces an operationally simple thresholding rule that is grounded in model derived scores rather than heuristic match scores. However, near-duplicates and covers can preserve sufficient structure to induce high-confidence false positives. These failure modes define an important limitation and motivate further study and targeted stress tests on near-duplicates.

A practical challenge in internalizing a dataset is update cost. Unlike retrieval based systems that ingest new tracks simply by inserting items into an external index, direct recognition requires updating model parameters without forgetting previously learned tracks. We evaluate incremental dataset growth under fine-tuning, retraining and continual-learning updates. Naive fine-tuning tends to acquire new tracks rapidly but increases forgetting. Continual-learning strategies, especially rehearsal, reduce forgetting at lower cost than retraining, making scheduled maintenance plausible. More broadly, internalization shifts the burden of dataset growth to offline model maintenance. Explicit retrieval or hybrid designs therefore may remain preferable for datasets with high-frequency ingestion and tight update-latency budgets. Efficient updates are a general challenge for parameterized models and remain an active area of research.

Moreover, because the dataset is represented implicitly by finite model capacity and accessed through a tokenized identifier, error can grow with dataset size. An important direction for future work is to study scaling at much larger dataset sizes; including replacing flat identity tokens with hierarchical or compositional identifiers that factor identity into shared structure and mitigate interference as the dataset grows. Moreover, musical identity is not always canonical, exact-ID prediction can be complicated in cases that are perceptually close or legally ambiguous. Version-aware datasets and evaluation protocols are necessary to fully evaluate robustness.

Despite these limitations, our results suggest that explicit retrieval is not fundamental in the closed-set regime we study. During training, an audio dataset can be internalized into model parameters. This allows a query to be resolved by direct recognition, while remaining robust and efficient. The consequence is not merely improved speed or reduced storage, but a different computational posture. Direct recognition offers a different design axis for information systems where the boundary between memory and computation blurs, bringing search closer to human associative recognition where partial cues trigger recall without explicit enumeration. In this view, search is not an algorithm we run, but a capability acquired during training. We do not argue that generative models replace retrieval in general; rather, direct recognition is viable when identity is unambiguous and the answer set is fixed.

## Method

### Closed-set identification as dataset memorization

We study closed-set music identification over a fixed dataset of $N$ audio tracks. Each track is assigned a unique discrete tokenized identifier. Given a short query excerpt sampled from an unknown temporal offset within one of the dataset tracks, the system predicts the corresponding track identifier. Performance is reported as Top-1 identification accuracy.

Unlike open-world classification, the objective here is not to generalize to unseen identities, but to internalize a fixed and finite dataset. The task therefore requires the model to *memorize* the dataset itself. The model learns to store the mapping from audio excerpts to identifiers in its parameters and generalize only within identity — across time offsets and identity-preserving transformations — rather than across new classes.

Because the goal is dataset internalization, the identity set is fixed and fully observed during training. We do not perform an identity-level train/validation split. Instead, we evaluate whether the model has successfully internalized the dataset by sampling random excerpts from each track after training and measuring exact identifier recovery. Here, Top-1 accuracy directly reflects how completely the dataset has been internalized.

**Data sources and preprocessing**

Dataset construction

Music tracks are sampled from the Free Music Archive (FMA) full release [14]. For each dataset size $N \in \{100, 1{,}000, 5{,}000, 10{,}000, 25{,}000\}$, we select $N$ distinct tracks at random. Each list is fixed after construction and shared across all five seeded runs. All audio is converted to WAV using ffmpeg library, downmixed to mono by channel averaging, and resampled to 16 kHz. Unless stated otherwise, all processing operates on 16 kHz mono float32 waveforms. We do not enforce artist-level disjointness and do not filter near-duplicates, remasters, or alternate versions within FMA.

Curated version shift robustness set

To evaluate robustness to alternate versions that preserve musical identity but differ substantially from the reference waveform, we curate a separate dataset of 1,000 musical works. Each work contains three paired versions: an original reference recording, a radio edit, and a live version. We define a radio edit as a broadcast-oriented or commercially edited version of the same work, which may differ from the original through shortened sections, modified intros or outros, fades, censoring, remastering, voiceovers, or other production changes. We define a live version as a concert or live-performance recording of the same work, which may differ in tempo, arrangement, instrumentation, vocal delivery, audience noise, reverberation, and room acoustics. This dataset is used only for the robustness experiments in Table 4. For evaluation, radio-edit and live-version queries are mapped to the identifier of their paired original reference recording, so these conditions test work-level version robustness rather than exact recording-level matching.

**Dataset protocol, open-set negatives, and calibration**

Closed-set protocol

For closed-set identification, the entire dataset is used during training. We do not create train/validation/test splits over identities, as the objective is to memorize the dataset itself. For evaluation, we sample random excerpts from the same $N$ tracks and compute top-1 identification accuracy, i.e., whether the predicted track identifier exactly matches the ground-truth label. This tests whether the model can map time-shifted and optionally distorted short excerpts to the correct track, assessing both memorization and within-track robustness.

Out-of-dataset pool

Since copyright monitoring systems must also handle queries that are not part of the monitored dataset, we evaluate open-set operation, in which the system must either assign a dataset identifier or reject the query as out-of-dataset. For this evaluation, we construct an out-of-dataset pool by sampling an additional $N{=}1{,}000$ tracks uniformly at random from FMA, ensuring no file-level overlap with the in-dataset tracks. This pool is used exclusively for open-set evaluation and decision-threshold calibration.

Calibration for open-set thresholds

Open-set rejection requires threshold selection. We therefore build a calibration set consisting of query excerpts from both within-dataset ("seen") and out-of-dataset ("unseen") tracks. Calibration is used only to set decision thresholds and is not used for model optimization.

**Query generation**

Single-excerpt queries

For a query duration $\ell$, we uniformly sample a start offset at random over all feasible offsets in the track and extract an excerpt of length ($\ell$). Unless otherwise stated, each seeded run uses one independently sampled excerpt per track. No constraints are imposed on inference offsets; and excerpts may originate from any temporal position within the track.

Multi-excerpt queries
For open-set evaluation and risk reduction, we additionally employ a streaming-style protocol in which each audio file produces six contiguous excerpts. Specifically, we sample a single start offset and extract six consecutive, non-overlapping segments from that position. Each segment is processed independently, and their predictions are aggregated to form a final decision.

Randomness and reproducibility
Experiments are repeated over five pre-specified random seeds: 1337, 1338, 1339, 1340, and 1341. For each seeded run, the seed is applied to the Python, NumPy, and PyTorch random number generators and controls model initialization, minibatch/window sampling, augmentation selection, and query-offset sampling. Dataset manifests are fixed after construction and shared across seeds. Query excerpts are sampled on the fly rather than stored explicitly. Reported results are averaged across the five seeded runs.

**Time–frequency representation**

All audio is processed at 16 kHz. Log-mel features are computed using torchaudio with:

- STFT: n_fft = 1024, win_length = 1024, hop_length = 320
- Window: Hann (torchaudio default)
- Spectrogram: power spectrogram (power = 2)
- Mel filterbank: n_mels = 64, torchaudio defaults f_min = 0 Hz, f_max = 8000 Hz
- Log compression: power-to-dB via AmplitudeToDB
- No normalization (no per-example or dataset-level normalization)

No per-example or dataset-level normalization is applied. In the neural model pipeline, feature extraction is performed on GPU for efficiency; the transformation itself is identical.

**Patch tokens and effective duration**

Each query is represented as a sequence of non-overlapping time–frequency patches extracted from the log-mel spectrogram. With a 20 ms hop, each patch advances by 8 frames, corresponding to a 160 ms patch stride in spectrogram time. Because the STFT window length is 1024 samples, or 64 ms at 16 kHz, the waveform support of P patches is $(8P-1)\times 20$ ms+64 ms. Thus, for $\ell = 1$ s, P = 7 patches, corresponding to 1.12 s of spectrogram stride and 1.164 s of waveform support.

**Efficient training pipeline**

Training time is dominated by window sampling, log-mel computation, and filtering near-silent segments. To reduce overhead, we cache valid (non-silent) window start indices per track and construct a GPU-resident pool of training windows.

Silence criterion
A waveform window is considered non-silent if: $mean(|x| > 10^{-4})$. This lightweight criterion is used during training to filter out low-energy segments.

Cached valid start indices
For each WAV file, we enumerate candidate window start indices on a fixed temporal grid (stride set by the training configuration), evaluate the silence criterion, and store the list of valid starts as a per-track JSON file. Cache validity is keyed by sample rate, window length in samples, stride in samples, silence threshold. Any change to these parameters invalidates the cache and triggers regeneration. Silence caching is used exclusively during training for efficiency. Inference excerpts are sampled uniformly at random and do not rely on cached indices.

GPU window pool
For each minibatch of songs, waveform windows are loaded from cached valid start positions and optionally augmented (time offset, additive noise, impulse response). All waveform segments are stacked into a single GPU batch, transformed to log-mel features, and segmented into non-overlapping time patches (flattened into token vectors).

We enforce a fixed token length per experiment ($P$ patches). Windows that produce fewer than $P$ patches are discarded; longer windows are truncated to the first $P$ patches.

This produces two GPU-resident tensors:

- windows_pool: shape $(M, P, D)$, where $D$ = n_mels × patch_t

- id_pool: shape ($M$), containing the corresponding track identifiers

Training minibatches are sampled directly from this pool, avoiding repeated CPU–GPU transfers and redundant feature computation.

**Training-time augmentation for direct recognition**

We adapt the waveform-level augmentation categories used in neural audio fingerprinting for direct recognition [3], and additionally apply SpecAugment-style time–frequency masking to the log-mel representation [13].

Time offset ($p = 0.5$): A random temporal shift of up to ±200 ms is applied by sampling a slightly longer waveform segment and cropping from a randomly shifted start position within this range.

Background noise ($p = 0.5$): Additive noise is mixed at a randomly sampled SNR in [0, 10] dB. Noise signals are resampled to match the target sample rate if necessary and length-matched via repetition (if shorter) or random cropping (if longer). Noise is RMS-scaled ($\varepsilon = 10^{-9}$ for numerical stability) before mixing, and the resulting waveform is clipped to [−1, 1].

Impulse response ($p = 0.5$): Room impulse responses (IRs) are applied via FFT-based convolution. IRs are RMS-normalized prior to convolution, and up to two IRs may be applied sequentially. The output waveform is clipped to [−1, 1].

Spectrogram masking ($p = 0.5$): After log-mel extraction, SpecAugment-style masking is applied, consisting of a rectangular cutout combined with additional time and frequency masks. Mask extents are sampled uniformly between 10% and 50% of the corresponding axis length.

Post-augmentation safeguard: If augmentation results in silence, defined as $mean(|x| \leq 10^{-4})$ the augmented waveform is discarded and the clean waveform is used instead.

**Identifier encoding**

Track identifiers are parsed as integers from the filename stem. Each identifier is encoded as four tokens over a 256-symbol vocabulary (byte values 0–255) using little-endian byte order. Specifically, for an integer identifier $id$, the token sequence is:

$$y = \left[(\mathrm{id} \gg 0)\,\&\,255,\ (\mathrm{id} \gg 8)\,\&\,255,\ (\mathrm{id} \gg 16)\,\&\,255,\ (\mathrm{id} \gg 24)\,\&\,255\right]$$

No EOS or PAD tokens are used; the model employs a learned BOS embedding for decoding.

**Direct recognition model**

Architecture

We use a decoder-only Transformer with causal self-attention [21]. The model consumes $P$ patch tokens (projected to d_model) as context and autoregressively generates the 4 identifier tokens.

We evaluate three model sizes.

The Small model uses:

- Layers: 3
- d_model = 256
- Heads: 8
- FFN dimension: 1024
- Dropout: 0.0
- Positional encoding: RoPE (base 10,000) [22].

The model has 2,764,800 parameters and a 11 MB FP32 checkpoint on disk (including serialization overhead).

The Medium model uses:

- Layers: 3
- d_model = 512
- Heads: 8
- FFN dimension: 1024
- Dropout: 0.0
- Positional encoding: RoPE (base 10,000)

The model has 9,853,824 parameters and a 38 MB FP32 checkpoint on disk (including serialization overhead).

The Large model uses:

- Layers: 6
- d_model = 512
- Heads: 8
- FFN dimension: 1024
- Dropout: 0.0
- Positional encoding: RoPE (base 10,000)

The model has 19,310,976 parameters and a 74 MB FP32 checkpoint on disk (including serialization overhead).

Computation settings
We enable TF32 matrix multiplications on CUDA (allow_tf32=True, set_float32_matmul_precision("high")) to accelerate training while keeping FP32 training.

**Training procedure and checkpoint selection**

Training is performed separately for each dataset size $N$ and query duration $\ell$. For a fixed $\ell$, the input length is fixed to $P$ patches. We use two batching levels:

Song batch (window pool construction)
SONG_BATCH_SIZE = 32 for ($\ell \in \{1,2,3\}$) s
SONG_BATCH_SIZE = 16 for ($\ell \in \{5,10\}$) s

Window batch (optimization step)
SEGMENT_BATCH_SIZE = 1024 windows/step for ($\ell \in \{1,2,3\}$) s
SEGMENT_BATCH_SIZE = 512 windows/step for ($\ell \in \{5,10\}$) s.

Optimization uses AdamW with learning rate $3x10^{-4}$ and gradient clipping (maximum norm 1.0) [23]. PyTorch default AdamW settings are used. Training continues until near-complete dataset memorization is achieved, which typically occurs between epochs 150 and 200 depending on the model/dataset size. The model is trained with teacher forcing to minimize the mean negative log-likelihood of the four ground-truth identifier tokens.

Checkpoint selection
Because the objective is to internalize a fixed dataset, we select the checkpoint with the highest training Top-1 accuracy. Training accuracy is computed after each epoch over sampled windows. For each ($N$, $\ell$) configuration, this procedure selects the model that most completely memorizes the target dataset under the closed-set objective.

**Inference**

Closed-set identification
For a query excerpt, we compute patch tokens and decode four identifier bytes using greedy decoding (fixed length 4). The predicted integer is reconstructed from the decoded bytes and compared to ground truth.

Multi-segment aggregation and confidence
Under the multi-segment protocol, each query file produces $k$ segment-level predictions. For each segment, the model outputs a predicted track identifier. From these $k$ segments, we compute:

- Majority-vote: The *id* appearing most frequently among the $k$ segment-level predictions.
- Agreement: The fraction of segments whose predicted identifier matches majority-vote *id*.

Open-set rejection
A query is accepted as in-dataset if and only if:

$$\text{agreement} \geq \tau_a$$

Otherwise, the query is rejected as out-of-dataset. The threshold ($\tau_a$) is selected using a calibration set containing both within-dataset and out-of-dataset queries. We sweep candidate threshold and select an operating point that balances correct acceptance of within-dataset queries against false acceptance of out-of-dataset queries. The threshold is global and fixed after calibration for all reported experiments.

**Fingerprinting baseline**

We compare against the open-source Dejavu fingerprinting system using a PostgreSQL backend [15]. Unless stated otherwise, Dejavu's default fingerprint parameters are used.

We evaluate two database conditions:

1. Clean DB: fingerprints extracted from clean dataset tracks.
2. Augmented DB: fingerprints extracted from dataset tracks after applying only waveform-level background noise and impulse response convolution (time offset and spectrogram masking are not used for fingerprinting).

Noise files are split by file: 80% used for augmented DB construction and 20% held out for query-time noise, ensuring evaluation noise differs from DB augmentation noise. At query time, Dejavu's top match is treated as the predicted identifier. For open-set rejection, we threshold Dejavu confidence outputs (e.g., input_confidence) using the same calibration protocol as for the neural method.

**Metrics**

Closed-set performance is evaluated using Top-1 identification accuracy, defined as the proportion of query excerpts for which the predicted track identifier exactly matches the ground-truth dataset identifier. Because each query corresponds to a single valid identity, Top-1 accuracy directly measures the degree to which the dataset has been successfully internalized. For open-set evaluation, we augment the test set with out-of-dataset queries and assess both identification and rejection performance. We report overall accuracy, precision, recall, F1 score, false positive rate (FPR), false negative rate (FNR), and balanced accuracy.

**Systems measurements: latency and storage**

All experiments were run on Linux (KVM virtualized) with an Intel Xeon CPU @ 2.20 GHz (6 cores / 12 threads), ~47 GiB RAM, and an NVIDIA L4 GPU (23 GB VRAM), with NVIDIA driver 550.90.07 and CUDA 12.4. Software versions: Python 3.10.17; PyTorch 2.9.1+cu128; torchaudio 2.9.1+cu128; NumPy 2.2.6; soundfile 0.13.1; PostgreSQL 10.7; Dejavu commit as above.

Latency is measured per query using time.perf_counter(). For GPU timing, we use torch.cuda.synchronize() before and after inference. Timing includes preprocessing (mel + dB), patchification, and decoding at batch size 1.

Fingerprinting storage is the on-disk PostgreSQL database size (reported separately for clean and augmented DBs). Direct recognition storage is the saved PyTorch checkpoint size. Raw audio storage is excluded.

**Use of Large Language Models**

The authors used ChatGPT and Grammarly for language editing (grammar, readability, phrasing) of portions of this manuscript. All suggested edits were reviewed and applied manually [24, 25].

**Reproducibility and availability**

All training and evaluation procedures are repeated over five pre-specified random seeds. Reported aggregate results are averaged across the five runs. We will release code, dataset manifests, and scripts to reproduce preprocessing, training, evaluation, and fingerprint database construction, subject to any licensing constraints of the underlying audio sources.

**Data Availability Statement**

The Free Music Archive data analysed during the current study are publicly available at https://doi.org/10.24432/C5HW28. The datasets generated during the current study are available from the corresponding author on reasonable request.

**Author Contributions Statement**

M.T.H. and A.H. contributed equally to this work. M.T.H. and A.H. performed the experiments and analyses, interpreted the results, and wrote the initial draft of the manuscript. G.X. conceived and designed the study, supervised the project, provided scientific guidance, and contributed to interpretation of the findings. All authors reviewed, edited, and approved the final manuscript.

**Funding Declaration**

The authors declare that this research work was conducted as part of the authors' university research group activities, and no funding was received for the research, authorship, or publication of this work.

**Clinical Trial Number: not applicable.**

**Consent to Participate declaration: not applicable.**

**Consent to Publish declaration: not applicable.**

**Ethics declaration: not applicable.**